\documentclass[11pt]{article}
\usepackage{amsmath}
\usepackage{amssymb}
\usepackage{bbm}
\usepackage{epsfig}
\allowdisplaybreaks[4]
\begin{document}
\title{One-loop charge and colour breaking associated with the top Yukawa
coupling}
\author{P.M. Ferreira~\footnote{ferreira@cii.fc.ul.pt} \\ Centro de F\'{\i}sica 
Te\'orica e Computacional, \\ Universidade de Lisboa, Av. Prof. Gama Pinto, 2, 
1649-003 Lisboa, Portugal }
\date{June, 2004} 
\maketitle
\noindent
{\bf Abstract.} We calculate the one-loop contributions to the MSSM effective 
potential when the scalar top fields have non-zero vacuum expectation values
and study their impact on charge and colour breaking bounds. 
\vspace{1cm}

Supersymmetry is presently the best theoretical extension of the Standard Model 
(SM) of particle physics. Its popularity derives from the fact it solves both 
the hierarchy and naturalness problems, stabilizing the weak scale by keeping 
huge quadratic divergences under control. Other incentives for supersymmetry 
include it making gauge unification more plausible and the tantalizing prospect 
of supergravity theories aiding in the formulation of a quantum theory of 
gravity. The most unpleasant aspect of supersymmetry is the amount of free 
parameters it introduces: assuming soft supersymmetry breaking we would have, in
addition to the $\sim$ 20 free parameters of the SM, about 100 new input values.
Charge and/or colour breaking (CCB) bounds have been used widely to try to limit
the size of that parameter space. Its foundations are simple: in the Minimal 
Supersymmetric Standard Model (MSSM) electroweak gauge symmetry is broken when 
the neutral components of the two Higgs fields acquire non-zero vacuum 
expectation values. As these fields are also colourless the remaining gauge 
symmetry is $SU(3)_C\times U(1)_{em}$, which is according to the experimental 
observation that we live in a world where electric charge and colour symmetries 
are unbroken. As we want to preserve angular momentum conservation, only fields 
of spin zero may have a non-zero vev. In the SM this is hardly a problem, since
there is but one scalar field in the theory. 

In the MSSM we have a plethora of scalars, most of them carrying colour, charge 
or leptonic number. It is then theoretically possible that, should some of these
fields have vevs, the MSSM potential develop a minimum deeper than the usual 
one, where only $H_1^0$ and $H_2^0$ have vevs. In that case the deeper minimum 
would surely violate charge and/or colour conservation. This appealing idea was 
first introduced in~\cite{fre} and used to impose bounds on the supersymmetric 
input values: if a particular combination of them causes the appearance of a CCB
minimum, we can exclude it on experimental grounds. A great deal of work was 
done on CCB bounds~\cite{eve}, of particular importance that of Gamberini et 
al.~\cite{gam}, showing the relevance of the one-loop corrections to the 
effective potential for gauge symmetry breaking, and the need to choose the 
appropriate renormalisation scale $M$ to better take them into account. The 
choice of $M$ became of fundamental importance in evaluating CCB bounds, so as 
to avoid an excessive constriction of the parameter space. A thorough review of 
the matter may be found in ref.~\cite{cas} and recent use of CCB bounds in 
phenomenological studies in refs.~\cite{phe}. A common point in all of these 
analysis was that only the tree-level potential was studied. It was argued in 
ref.~\cite{gam} that a careful choice of renormalisation scale would correctly 
reproduce the effects of the one-loop minimisation of the potential, both for 
the ``real" minimum as for the CCB one. This led to comparing the value of the 
potential at the two minima at two different renormalisation scales. 

Partial one-loop contributions to the CCB bounds were considered in 
ref.~\cite{one} using renormalisation group arguments. Recently a full one-loop 
charge breaking effective potential was calculated for the case where the scalar
fields $\tilde{\tau}_L$ and $\tilde{\tau}_R$, not simply $H_1^0$ and $H_2^0$, 
acquire a non-zero
vev~\cite{ccb}. At the same time it was argued that one should compare the 
values of the ``real" potential and the CCB one at the same renormalisation 
scale due to the presence, in the effective potential, of field-independent 
terms. The minimisation of this one-loop CCB potential showed the Gamberini et 
al. strategy was not working for the CCB one-loop contributions~\cite{min} and 
raised serious questions about the perturbative 
believability of bounds derived from this particular CCB direction. This would 
seem to be closely related to the smallness of the tau Yukawa coupling 
associated with this pattern of CCB, which generated typical CCB sparticle 
masses of the order of tens of TeV. CCB associated with a larger Yukawa should 
not fall victim to the same perturbative problems, and the only Yukawa coupling 
that satisfies those conditions is the top one~\footnote{The bottom Yukawa is 
larger than the tau one, but still of the same order of magnitude.}. In this 
paper we undertake the calculation of the full one-loop effective potential for 
the CCB case where the scalar fields $\tilde{t}_L$ and $\tilde{t}_R$, $H_1^0$ 
and $H_2^0$, 
acquire non-zero vevs. In section~\eqref{sec:der} we review the model we will be
working in and our conventions. We also discuss our strategy for CCB bounds and 
review the results of refs.~\cite{ccb,min}. The CCB mass matrices are introduced
in section~\eqref{sec:mas} and the minimisation of the one-loop potential, and 
its results, is shown in section~\eqref{sec:min}. We conclude with an analysis 
of the results and their relevance for CCB bounds. 

\section{Charge and colour breaking in the MSSM}
\label{sec:der}

We follow most of the conventions of ref.~\cite{bbo} (except for the sign of the
$\mu$ parameter). Because the Yukawa couplings of the first and second 
generations are so small compared to those of the third, we set them to zero.
The MSSM superpotential is thus given by
\begin{equation}
W \; = \; \lambda_t H_2\,Q\,t_R \;+ \; \lambda_b H_1\,Q\,b_R \; +\; \lambda_\tau
H_1\,L\,\tau_R \; +\; \mu H_2\,H_1 \;\;\; ,
\end{equation}
with $SU(2)$ doublets $H_1 = (H_1^0 \,,\, H_1^-)$, $H_2 = (H_2^+ \,,\, H_2^0)$, 
$Q = (t_L \,,\, b_L)$, $L = (\nu_L \,,\, \tau_L)$ and singlets $t_R$, $b_R$ and 
$\tau_R$. The MSSM tree-level effective potential is the sum of the $F$ and $D$ 
terms and the soft supersymmetry breaking potential. Its full expression may be
found, for instance, in references~\cite{ccb} and~\cite{min}. At the 
renormalisation scale $M$ the one-loop contributions to the potential are
\begin{equation}
\Delta V_1 \, =\, \sum_\alpha \, \frac{n_\alpha}{64\pi^2}\,M_\alpha^4\, \left(\,
\log \frac{M_\alpha^2}{M^2}\, - \, \frac{3}{2}\,\right) \;\;\; ,
\label{eq:dv1}
\end{equation}
where the $M_\alpha$ are the (tree-level) masses of each particle of spin
$s_\alpha$ and $n_\alpha = (-1)^{2s_\alpha}\,(2s_\alpha +1)\,C_\alpha\,
Q_\alpha$, $C_\alpha$ being the number of colour degrees of freedom of each 
particle and $Q_\alpha$ counting its particle/anti-particle states. If only 
$H_1^0$ and $H_2^0$ acquire vevs $v_1/\sqrt{2}$, $v_2/\sqrt{2}$, the negative 
contributions to the tree-level potential come from the $-\,B\,\mu\,v_1\,v_2$ 
term (and usually from the $m_{H_2}^2\,v_2^2$ term as well). So that CCB occurs 
extra negative contributions are necessary - these come from terms cubic in the 
vevs, from the soft and F-term potentials. There are many possible choices of 
CCB directions, but one must remember that along with negative trilinear 
contributions to $V_0$ come potentially large positive quadratic and quartic
terms as well. In refs.~\cite{ccb,min} we studied in detail the case where the 
fields $\tilde{\tau}_L$, $\tilde{\tau}_R$ had non-zero vevs. This particular 
direction, being associated with the tau Yukawa coupling, was in principle quite
favourable to CCB - not only are $m_L$, $m_\tau$ usually the smallest of the 
soft masses but the size of $\lambda_\tau$ should reduce the magnitude of the 
F-term contributions to the potential. Finally, as vevs at such a CCB minimum 
should be of the order of $v \sim g_2\,A_\tau/\lambda_\tau$, the resulting 
potential, if negative, ought to be much deeper than the ``real" minimum. The 
CCB direction we propose to study in this article - the scalar fields 
$\tilde{t}_L$ and $\tilde{t}_R$ acquiring vevs - has none of these advantages: 
$m_Q$, $m_t$ are usually the largest of the soft masses; the top Yukawa being of
order 1, the F-term contributions will be large; and the CCB potential, if 
negative, is not at all guaranteed to be deeper than the ``real" one. 
Nonetheless, as explained in the introduction, we hope this CCB direction is not
afflicted by the perturbative problems found in ref.~\cite{min}. Also, the 
arguments against top Yukawa CCB rely on an intuitive analysis of the tree-level
potential. Such analysis, however, is not possible in the case of the one-loop 
contributions, which are very complex. Specifying notation, we must remember 
that $\tilde{t}_L$, $\tilde{t}_R$ are in the $3$, $\bar{3}$ representations of 
$SU(3)$ respectively - each field has therefore three colour degrees of freedom.
To limit the size of the $SU(3)$ D-terms it is best to choose both vevs having 
the same $SU(3)$ index - to simplify calculations we chose the third colour 
index~\footnote{Identical results would be obtained for the colour indices $(1)$
or $(2)$.}. We emphasize that this choice is {\em not} the most general CCB case
associated with the top Yukawa, merely the one we expect will produce more 
interesting minima. So, let us consider that $\{\tilde{t}_L^{(3)}\,,\, 
\tilde{t}_R^{(3)}\}$ have vevs $\{q/\sqrt{2}\,,\,t/\sqrt{2}\}$ and $\{H_1^0\,,\,
H_2^0\}$ have vevs $\{v_1/\sqrt{2}\,,\,v_2/\sqrt{2}\}$ as usual. The tree-level 
potential then becomes
\begin{eqnarray}
V_0 & = & \frac{\lambda_t^2}{4}\,\left[v_2^2\,(q^2+t^2)\,+\,q^2\,t^2\right]\;-\;
\frac{\lambda_t}{\sqrt{2}}\,(A_t\,v_2\,-\,\mu\,v_1)\,q\,t\;+\;\frac{1}{2}\,(
m_1^2\,v_1^2\,+\,m_2^2\,v_2^2\,+\,m_Q^2\,q^2\, \nonumber \\
 & & +\;m_t^2\,t^2)\;-\;B\,\mu\,v_1\, v_2 \;+\; \frac{g^{\prime 2}}{32}\,
\left(v_2^2-v_1^2+ \frac{1}{3}\,q^2-\frac{4}{3}\,t^2\right)^2\;+\;
\frac{g_2^2}{32}\,(v_2^2-v_1^2- q^2)^2 \nonumber \\
 & & +\;\frac{g_3^2}{24}\,(q^2-t^2)^2 \;,
\label{eq:vc}
\end{eqnarray}
with $m_1^2\,=\,m_{H_1}^2\,+\,\mu^2$ and $m_2^2\,=\,m_{H_2}^2\,+\,\mu^2$. This
potential being a 4-variable function analytical studies of CCB are possible 
only in simplified cases. The usual strategy~\cite{eve} considers only the 
tree-level value of the potential and the tree-level derived vevs: the latter 
simplification is based on ref.~\cite{gam} where it was showed that for the 
``real" MSSM potential, by choosing a renormalisation scale $M$ of the order of 
the largest mass present in $\Delta V_1$, the tree-level vevs were a good 
approximation to the one-loop ones - that is to say, in this range of $M$ the 
one-loop contributions to the minimisation conditions, given by
\begin{equation}
\sum_\alpha \, \frac{n_\alpha}{32\pi^2}\,M_\alpha^2\,\frac{\partial M_\alpha^2}{
\partial v_i}\, \left(\, \log \frac{M_\alpha^2}{M^2}\, - \, 1\,\right) \;\;\; ,
\label{eq:der}
\end{equation}
are not significative. Notice, however, that this does not necessarily imply 
that the contributions $\Delta V_1$ will be negligible. In fact, even in the 
usual MSSM calculations with an appropriate mass scale, the value of the 
one-loop potential is {\em positive} whereas the tree-level one is {\em 
negative}. This is not a problem since the value of the potential, CCB bounds 
excluded, has no bearing on the phenomenological aspects of the model. But it 
shows that even if an adequate choice of scale may render insignificant the 
{\em derivatives} of the one-loop contributions to the potential, that same 
choice does not imply that $\Delta V_1$ itself is negligible. Since the typical 
mass is usually different in the ``real" and CCB potentials this leads to both 
being compared at different renormalisation scales. It was this point that led 
us to argue against this procedure in ref.~\cite{ccb}, based on the fact that 
the sum $V_0 + \Delta V_1$ is not renormalisation group (RG) invariant. Instead,
the complete RG invariant effective potential is given by~\cite{ford}
\begin{equation}
V(M,\lambda_i,\phi_j) \; =\; \Omega(M,\lambda_i) \;+\; V_0(\lambda_i,\phi_j) \;
+\; \hbar \,\Delta V_1 (M,\lambda_i,\phi_j)\; +\; O(\hbar^2) \;\;\; ,
\label{eq:om}
\end{equation}
where $\lambda_i$ stands for the couplings and masses of the theory and $\phi_j$
for its fields. The field-independent function $\Omega$, implicitly or 
explicitly, depends on the renormalisation scale $M$. The only difference 
between the CCB and ``real" potentials being the different set of values some of
the fields $\phi_j$ have, $\Omega$ is the same in both cases - so, if we compare
$V^{MSSM}$ and $V^{CCB}$ at different scales the contributions from $\Omega$ 
will not be considered correctly. But comparing potentials for the same value of
$M$ means that at least in one case the one-loop contributions to the vevs will
have to be included. By consistency, if $\hbar$ contributions to the vevs are 
being taken, $\hbar$ contributions to $V$, that is, $\Delta V_1$, must also be
considered. In ref.~\cite{min} we found an even stronger argument for studying
one-loop CCB potentials and their minimisation: eq.~\eqref{eq:om} implies that
$d(V_0+\Delta V_1)^{CCB}/d M \;=\; d(V_0+\Delta V_1)^{MSSM}/d M$, up to two-loop
effects. In other words, the two potentials must run with the renormalisation 
scale parallel to one another. 
In~\cite{min} we found that the CCB vevs were so large that the one-loop CCB
effective potential was {\em not} RG invariant. Perturbation theory had broken
down and the typically small two-loop effects had become quite large. As a 
consequence, the condition ``$V^{CCB} < V^{MSSM}$" was renormalisation scale
dependent and as such bounds thereof derived not reliable. We expect this 
will not happen for the top-associated CCB direction, as the typical vevs should
be smaller. 

In short, our CCB strategy will be to calculate the one-loop CCB potential,
obtain from its minimisation the one-loop vevs $\{v_1\,,\,v_2\,,\,q\,,\,t\}$
and compare it to the one-loop minimised ``real" one-loop potential, at the
same renormalisation scale. For comparison, we will also perform tree-level 
minimisations of the potentials (both MSSM and CCB) and compare their values - 
even though, as follows from the discussion above, this tree-level procedure is
misleading and may induce errors. For the one-loop calculations we will  
need the CCB masses contributing to $\Delta V_1$. The complication that arises, 
as in ref.~\cite{ccb}, is the fact
that the vevs $q$ and $t$ cause the mixing between charged/neutral and 
coloured/colourless fields in the theory. For instance, the trilinear terms in 
the soft potential cause mixing between $\{H_1^0\,,\,H_2^0\,,\,\tilde{t}_L^{(3)}
\,,\,\tilde{t}_R^{(3)}\}$ - and further mixing between them arises from the F 
and D-terms. Likewise, similar mixing occurs between $\{H_1^-\,,\,H_2^+\,,\,
\tilde{b}_L^{(3)}\,,\,\tilde{b}_R^{(3)}\}$. The charged, pseudoscalar and 
CP-even Higgses will therefore have $4 \times 4$ mass matrices. Because colour 
symmetry has been broken, particles carrying colour indices $\{(1)\,,\,(2)\}$ 
will have different masses from those with colour index $(3)$ - the $\{(1)\,,\,
(2)\}$ squarks have mass matrices very similar to those of the non-CCB case. The
existence of vevs carrying colour degrees of freedom gives mass and electric 
charge to four gluons, three others remaining massless. The eighth gluon remains
neutral but becomes also massive by mixing with the $B_\mu$ and $W^3_\mu$ 
fields. For the fermions the mixing is even more extensive: the charginos become
a mixture of charged $SU(2)$ gauginos, the fermionic partners of the charged 
Higgses and the $(3)$ component of the bottom quark. The case of the neutralinos
is even more complex: a $7 \times 7$ mass matrix originates from the mix of 
neutral $U(1)$ and $SU(2)$ gauginos, fermionic partners of the neutral Higgses, 
$(3)$ components of the top quark and the eighth gluino. The gauge interaction 
term between quarks, scalar quarks and gluinos, $-i\,g_3\, \phi^\dagger\,
\lambda_i\,\tilde{\phi}\,\tilde{G}^i/\sqrt{2}$, also causes mixing between the 
$\{(1)\,,\,(2)\}$ quark components and the $\tilde{G}^{4\ldots 7}$ gluinos - the
results are two $4 \times 4$ identical mass matrices. The mass of the gluinos 
$\tilde{G}^{1\ldots 3}$ remains unchanged, $M_3$. We present the CCB mass 
matrices in the next section.

\section{Mass matrices}
\label{sec:mas}

Let us define the coefficients $G$ as 
\begin{equation}
G_1\;=\; v_2^2\,-\,v_1^2\,+\,\frac{1}{3} q^2\,-\,\frac{4}{3} t^2 \;\;,\;\;
G_2 \;=\; v_2^2\,-\,v_1^2\,-\,q^2 \;\;,\;\; 
G_3 \;=\; q^2\,-\,t^2 \; . 
\end{equation}
We now list the masses of the MSSM  when $q$ and $t$ are non-zero. 
\begin{itemize}
\item First and second generation sleptons and sneutrinos ($n_1 = n_2 = 
n_{\tilde{\nu}_e} = 2 \times 2$)~\footnote{ We considered degenerate first and 
second generation sparticles, but these results are trivially generalised.}:
\begin{align}
M^2_{\tilde{e}_1} & = m_N^2 \;-\;\frac{{g^\prime}^2}{8}\,G_1\,+\,\frac{g_2^2}{8}
\,G_2 & M^2_{\tilde{e}_2} & =  m_e^2\,+\,\frac{{g^\prime}^2}{4}\,G_1\nonumber \\
M^2_{\tilde{\nu}_e} & = m_N^2 \,-\,\frac{{g^\prime}^2}{8}\,G_1\,-\, 
\frac{g_2^2}{8}\,G_2 & & \;\;\; .
\end{align}
\item First and second generation squarks, colour indices $\{1\,,\, 2\}$ ($n_1 =
n_2 = 2 \times 4$, for both up and down type squarks):
\begin{align}
{M^2_{\tilde{u}_1}}^{(1,2)} &= m_R^2 \,+\, \frac{{g^\prime}^2}{24}\,G_1\,-\,
\frac{g_2^2}{8}\,G_2\,-\,\frac{g_3^2}{12}\,G_3 & {M^2_{\tilde{u}_2}}^{(1,2)} &= 
m_u^2 \,-\,\frac{{g^\prime}^2}{6}\,G_1\,+\,\frac{g_3^2}{12}\,G_3 \nonumber \\
{M^2_{\tilde{d}_1}}^{(1,2)} &= m_R^2 \,+\,\frac{{g^\prime}^2}{24}\,G_1\,+\,
\frac{g_2^2}{8}\,G_2\,-\,\frac{g_3^2}{12}\,G_3 & {M^2_{\tilde{d}_2}}^{(1,2)} &= 
m_d^2 \,+\,\frac{{g^\prime}^2}{12}\,G_1\,+\,\frac{g_3^2}{12}\,G_3 \;\;\; . 
\end{align}
\item First and second generation squarks, colour index $3$ ($n_1 = n_2 = 2 
\times 2$, for both up and down type squarks):
\begin{align}
{M^2_{\tilde{u}_1}}^{(3)} &= m_R^2 \,+\, \frac{{g^\prime}^2}{24}\,G_1\,-\, 
\frac{g_2^2}{8}\,G_2\,+\,\frac{g_3^2}{6}\,G_3 & {M^2_{\tilde{u}_2}}^{(3)} &= 
m_u^2 \,-\,\frac{{g^\prime}^2}{6}\,G_1\,-\,\frac{g_3^2}{6}\,G_3\nonumber \\
{M^2_{\tilde{d}_1}}^{(3)} &= m_R^2 \,+\,\frac{{g^\prime}^2}{24}\,G_1\,+\, 
\frac{g_2^2}{8}\,G_2\,+\,\frac{g_3^2}{6}\,G_3 & {M^2_{\tilde{d}_2}}^{(3)} &=
m_d^2 \,+\,\frac{{g^\prime}^2}{12}\,G_1\,-\,\frac{g_3^2}{6}\,G_3 \;\;\; .
\end{align}
\item Tau sneutrino and sleptons ($n_{\tilde{\nu}_\tau} = n_1 = n_2 = 2$):
\begin{equation}
M^2_{\tilde{\nu}_\tau} \;=\; m_L^2 \,-\,\frac{{g^\prime}^2}{8}\,G_1\,-\,
\frac{g_2^2}{8}\,G_2 \;\;\;,\;\;\; 
[M^2_{\tilde{\tau}}]\; =\;\begin{pmatrix} a_{\tilde{\tau}} & b_{\tilde{\tau}}\\
b_{\tilde{\tau}} & c_{\tilde{\tau}}\end{pmatrix} \;\;\; ,
\end{equation}
with
\begin{align}
a_{\tilde{\tau}} &= m_L^2 \,+\, \frac{\lambda_\tau^2}{2}\,v_1^2\,-\, 
\frac{{g^\prime}^2}{8}\,G_1\,+\,\frac{g_2^2}{8}\,G_2 & b_{\tilde{\tau}} &=
\frac{\lambda_\tau}{\sqrt{2}}\,(\mu\, v_2 \, -\, A_\tau\, v_1) \nonumber \\
c_{\tilde{\tau}} &= m_\tau^2\,+\,\frac{\lambda_\tau^2}{2}\,v_1^2\,+\,
\frac{{g^\prime}^2}{4}\,G_1 \;\;\; . & & 
\end{align}
\item Charged ($n = 6$) and neutral ($n_1 = n_2 = 3$) electroweak gauge bosons:
\begin{equation}
M^2_W\; =  \; \frac{1}{4}\;g_2^2\;(v_1^2+v_2^2+q^2) \;\;\;,\;\;\;
[M^2_{G^0}]\; =\;\begin{pmatrix} a_{G^0} & b_{G^0} & d_{G^0} \\ b_{G^0} & 
c_{G^0} & e_{G^0} \\ d_{G^0} & e_{G^0} & f_{G^0}  
\end{pmatrix} \;\;\; ,
\end{equation}
with
\begin{align}
a_{G^0} &= \frac{g_2^2}{4}(v_1^2+v_2^2+q^2) & 
b_{G^0} &= -\frac{g_2\,g^\prime}{4}\left(v_1^2+v_2^2-\frac{1}{3}q^2\right)
\nonumber \\
c_{G^0} &= \frac{{g^\prime}^2}{4}\left(v_1^2+v_2^2+\frac{1}{9}q^2+
\frac{16}{9}t^2\right) &
d_{G^0} &= -\frac{g_2\,g_3}{2\sqrt{3}}q^2\nonumber \\
e_{G^0} &= -\frac{g^\prime\,g_3}{\sqrt{3}}\left(\frac{1}{6}q^2+\frac{2}{3}
t^2\right) &
f_{G^0} &= \frac{g_3^2}{3}(q^2+t^2) \;\;\; .
\end{align}
\end{itemize}
This matrix has one zero eigenvalue, corresponding to a ``photon" resulting from
gauge symmetry breaking~\footnote{The symmetry breaking we have chosen leaves
intact a $SU(2)\times U(1)$ gauge group, corresponding to an integer charge
quark theory. In this theory four gluons couple directly to the photon and as 
such possess electric charge. See reference~\cite{icq} for details.}. 
\begin{itemize}
\item Charged gluons ($n_1 = 4 \times 3$):
\begin{align}
M^2_{G^\pm} &= \frac{1}{4}\,g_3^2\,(q^2\;+\;t^2) \;\;\; .
\label{eq:mgl}
\end{align}
\item Top scalars, colour indices $\{1\,,\, 2\}$ ($n_1 = n_2 = 4$):
\begin{equation}
[{M^2_{\tilde{t}}}^{(1,2)}]\; =\;\begin{pmatrix} a_{\tilde{t}} & b_{\tilde{t}} 
\\
b_{\tilde{t}} & c_{\tilde{t}} \end{pmatrix} \;\;\; ,
\label{eq:mstop}
\end{equation}
with
\begin{align}
a_{\tilde{t}} &= m_Q^2\,+\,\frac{1}{2}\,\lambda_t^2\,v_2^2 \,+\,
\frac{{g^\prime}^2}{24}\,G_1\,-\,\frac{g_2^2}{8}\,G_2\,+\,\frac{g_3^2}{12}(3\,
q^2\,-\, G_3) \nonumber \\
b_{\tilde{t}} &= -\,\frac{\lambda_t}{\sqrt{2}}\,(A_t\,v_2\,-\,\mu\,v_1) \,+\,
\frac{1}{4}\,(2\,\lambda_t^2\,-\,g_3^2)\,q\,t \nonumber \\
c_{\tilde{t}} &= m_t^2 \,+\,\frac{1}{2}\,\lambda_t^2\,v_2^2 \,-\,
\frac{{g^\prime}^2}{6}\,G_1\,+\,\frac{g_3^2}{12} (3\,t^2\,+\,G_3) \;\;\; .  
\end{align}
\item Bottom scalars, colour indices $\{1\,,\, 2\}$ ($n_1 = n_2 = 4$):
\begin{equation}
[{M^2_{\tilde{b}}}^{(1,2)}]\; =\;\begin{pmatrix} a_{\tilde{b}} & b_{\tilde{b}} 
\\ b_{\tilde{b}} & c_{\tilde{b}} \end{pmatrix} \;\;\; ,
\end{equation}
with
\begin{align}
a_{\tilde{b}} &= m_Q^2\,+\,\frac{1}{2}\,\lambda_b^2\,v_1^2 \,+\,
\frac{{g^\prime}^2}{24}\,G_1\,+\,\frac{g_2^2}{8}\,G_2\,-\,\frac{g_3^2}{12} G_3
& b_{\tilde{b}} &= -\,\frac{\lambda_b}{\sqrt{2}}\,(A_b\,v_1\,-\,\mu\,v_2) 
\nonumber \\
c_{\tilde{b}} &= m_b^2 \,+\,\frac{1}{2}\,\lambda_b^2\,v_1^2 \,+\,
\frac{{g^\prime}^2}{12}\,G_1\,+\,\frac{g_3^2}{12} G_3 \;\;\; . & & 
\end{align}
\item Charged Higgs (mix between $H_1^-$, $H_2^+$, $\tilde{b}_L^{(3)}$ and 
$\tilde{b}_R^{(3)}$; $n_{1 \ldots 4} = 2$):
\begin{equation}
[M^2_{H^\pm}]\; =\;\begin{pmatrix} a_\pm & b_\pm & d_\pm & e_\pm \\ b_\pm & 
c_\pm & f_\pm & g_\pm \\ d_\pm & f_\pm & h_\pm & i_\pm \\ e_\pm & g_\pm & i_\pm 
& j_\pm \end{pmatrix} \;\;\; ,
\label{eq:mch}
\end{equation}
with
\begin{align}
a_\pm &= m_1^2\,+\, \frac{1}{2}\,\lambda_b^2\,q^2\,-\,\frac{{g^\prime}^2}{8}
\,G_1\,+\,\frac{g_2^2}{8}\,(v_1^2 + v_2^2 - q^2) \nonumber \\
 b_\pm & =B\,\mu\, + \, \frac{g_2^2}{4}\,v_1\,v_2 \nonumber \\
c_\pm & = m_2^2 \,+\, \frac{1}{2}\,\lambda_t^2\,t^2\,+\, \frac{{g^\prime}^2}{8}
\,G_1\,+\,\frac{g_2^2}{8}\,(v_2^2+v_1^2+q^2) \nonumber \\
d_\pm & = \frac{\lambda_t}{ \sqrt{2}}\,\mu\,t\,+\,\frac{1}{4}\,(g_2^2 - 2\,
\lambda_b^2)\,v_1\,q \nonumber \\
e_\pm &= \frac{1}{2}\,\lambda_b\,\lambda_t\,v_2\,t\,-\,\frac{\lambda_b}{
\sqrt{2}}\,A_b\,q \nonumber \\
f_\pm &= \frac{\lambda_t}{\sqrt{2}}\,A_t\,t\,+\,\frac{1}{4}\,(g_2^2 - 2\,
\lambda_t^2)\,v_2\,q \nonumber \\
g_\pm &= -\,\frac{\lambda_b}{\sqrt{2}}\,\mu\,q\,+\,\frac{1}{2}\,\lambda_b\,
\lambda_t\,v_1\,t \nonumber \\
h_\pm &= m_Q^2\,+\,\frac{1}{2}\,(\lambda_b^2\,v_1^2 + \lambda_t^2\,t^2)\,+\,
\frac{{g^\prime}^2}{24}\,G_1\,+\,\frac{g_2^2}{8}\,(v_2^2 - v_1^2 + q^2) \,+\,
\frac{g_3^2}{6}\,G_3 \nonumber \\
i_\pm &= \frac{\lambda_b}{\sqrt{2}}\,(A_b\,v_1\,-\,\mu\,v_2) \nonumber \\
j_\pm &= m_b^2 \,+\,\frac{1}{2}\,\lambda_b^2\,(v_1^2 + q^2) \,+\,
\frac{{g^\prime}^2}{12}\,G_1\,-\,\frac{g_3^2}{6} G_3 \;\;\; .
\end{align}
\item Pseudo scalars (mix between the imaginary parts of $H_1^0$, $H_2^0$, 
$\tilde{t}_L^{(3)}$ and $\tilde{t}_R^{(3)}$; $n_{1 \ldots 4} = 1$):
\begin{equation}
[M^2_{\bar{H}_0}]\; =\;\begin{pmatrix} a_{\bar{H}} & b_{\bar{H}} & d_{\bar{H}}
 & e_{\bar{H}} \\ b_{\bar{H}} & c_{\bar{H}} & f_{\bar{H}} & g_{\bar{H}} \\ 
d_{\bar{H}} & f_{\bar{H}} & h_{\bar{H}} & i_{\bar{H}} \\ e_{\bar{H}} & 
g_{\bar{H}} & i_{\bar{H}} & j_{\bar{H}} \end{pmatrix} \;\;\; ,
\label{eq:mps}
\end{equation}
with
\begin{align}
a_{\bar{H}} &= m_1^2\,-\,\frac{{g^\prime}^2}{8}\,G_1\, -\, \frac{g_2^2}{8}\,G_2
\nonumber \\
b_{\bar{H}} &= B\,\mu \nonumber \\
c_{\bar{H}} &= m_2^2\,+\,\frac{1}{2}\,\lambda_t^2\,(q^2 + t^2)\,+\,
\frac{{g^\prime}^2}{8}\,G_1\,+\,\frac{g_2^2}{8}\,G_2 \nonumber \\
d_{\bar{H}} &= \frac{\lambda_t}{\sqrt{2}}\,\mu\,t \nonumber \\
e_{\bar{H}} &= \frac{\lambda_t}{\sqrt{2}}\, \mu\,q \nonumber \\
f_{\bar{H}} &= \frac{\lambda_t}{\sqrt{2}}\,A_t\,t \nonumber \\
g_{\bar{H}} &= \frac{\lambda_t}{\sqrt{2}}\,A_t\,q \nonumber \\
h_{\bar{H}} &= m_Q^2 \,+\, \frac{1}{2}\,\lambda_t^2\,(v_2^2 + t^2)\,+\,
\frac{{g^\prime}^2}{24}\,G_1\,-\,\frac{g_2^2}{8}\,G_2\,+\,\frac{g_3^2}{6}\,G_3 
\nonumber \\
i_{\bar{H}} &= \frac{\lambda_t}{\sqrt{2}}\,(A_t\,v_2\,-\,\mu\,v_1) 
\nonumber \\
j_{\bar{H}} &= m_t^2\,+\,\frac{1}{2}\,\lambda_t^2\,(v_2^2 + q^2)\,-\,
\frac{{g^\prime}^2}{6}\,G_1\,-\,\frac{g_3^2}{6}\,G_3 \;\;\; .
\end{align}
\item Higgs scalars (mix between the real parts of $H_1^0$, $H_2^0$, 
$\tilde{t}_L^{(3)}$ and $\tilde{t}_R^{(3)}$; $n_{1 \ldots 4} = 1$):
\begin{equation}
[M^2_{H_0}]\; =\;\begin{pmatrix} a_H & b_H & d_H & e_H \\ b_H & c_H & f_H & 
g_H \\ d_H & f_H & h_H & i_H \\ e_H & g_H & i_H & j_H \end{pmatrix} \;\;\; ,
\end{equation}
with
\begin{align}
a_H &= m_1^2\,-\,\frac{{g^\prime}^2}{8}\,(G_1 - 2\,v_1^2)\, -\, \frac{g_2^2}{8}
\,(G_2 - 2\,v_1^2) \nonumber \\
b_H &= -\,B\,\mu\,-\,\frac{1}{4}\,({g^\prime}^2 + g_2^2)\, v_1\,v_2 \nonumber \\
c_H &= m_2^2\,+\,\frac{1}{2}\,\lambda_t^2\,(q^2 + t^2)\,+\,
\frac{{g^\prime}^2}{8}\,(G_1 + 2\,v_2^2)\,+\,\frac{g_2^2}{8}\,(G_2 + 2\,v_2^2)
\nonumber \\
d_H &= \frac{\lambda_t}{\sqrt{2}}\,\mu\,t\,+\,\frac{1}{12}\,(3\,g_2^2 - 
{g^\prime}^2)\,v_1\,q \nonumber \\
e_H &= \frac{\lambda_t}{\sqrt{2}}\, \mu\,q \,+\, \frac{1}{3}\,{g^\prime}^2\,v_1
\,t \nonumber \\
f_H &= \lambda_t^2\,v_2\,q\,-\,\frac{\lambda_t}{\sqrt{2}}\,A_t\,t \,+\,
\frac{1}{12}\,({g^\prime}^2 - 3\,g_2^2)\,v_2\,q \nonumber \\
g_H &= -\,\frac{\lambda_t}{\sqrt{2}}\,A_t\,q \,+\,\frac{1}{3}\,(3\,\lambda_t^2 -
{g^\prime}^2)\,v_2\,t \nonumber \\
h_H &= m_Q^2 \,+\, \frac{1}{2}\,\lambda_t^2\,(v_2^2 + 
t^2)\,+\,\frac{{g^\prime}^2}{24}\,\left(G_1\,+\,\frac{2}{3}\,q^2\right)\,-\,
\frac{g_2^2}{8}\,(G_2 - 2\,q^2)\,+\,\frac{g_3^2}{6}\,(G_3 + 2\,q^2) \nonumber \\
i_H &= -\,\frac{\lambda_t}{\sqrt{2}}\,(A_t\,v_2\,-\,\mu\,v_1)\,+\, \frac{1}{9}\,
(9\,\lambda_t^2 - {g^\prime}^2 - 3\,g_3^2)\,q\,t \nonumber \\
j_H &= m_t^2\,+\,\frac{1}{2}\,\lambda_t^2\,(v_2^2 + q^2)\,-\,\frac{{g^\prime}^2}
{6}\,\left(G_1 - \frac{8}{3}\,t^2\right)\,-\,\frac{g_3^2}{6}\,(G_3 - 2\,t^2) 
\;\;\; .
\end{align}
\item Tau lepton and bottom quark, colour indices $\{1 \,,\, 2\}$ ($n_\tau
 = -4$, $n_b = -2 \times 4$), $M_\tau \,=\, \lambda_\tau\,v_1/\sqrt{2}$, $M_b \,
=\, \lambda_b\,v_1/\sqrt{2}$. 
\item Charginos (mix between the fermionic partners of the charged Higgses 
$\tilde{H}_1^-$, $\tilde{H}_2^+$, the SU(2) gauginos $\Psi^+$, $\Psi^-$
and the colour index $3$ components of the bottom quark, $b_L^{(3)}$ and
$b_R^{(3)}$; $n_{1 \ldots 6} = -2$):
\begin{equation}
[M_{\chi^\pm}]\; =\;\,\begin{pmatrix} 0 & 0 & M_2 & -
\frac{g_2}{\sqrt{2}} \,v_1 & 0 & 0 \vspace{0.2cm} \\ 0 & 0 & 
\frac{g_2}{\sqrt{2}} \, v_2 & -\mu & \frac{\lambda_t}{\sqrt{2}} 
\,t & 0 \vspace{0.2cm} \\ M_2 & \frac{g_2}{\sqrt{2}}\, v_2 & 0 & 
0 & \frac{g_2}{\sqrt{2}} \,q & 0 \vspace{0.2cm} \\ -
\frac{g_2}{\sqrt{2}}\, v_1 & -\mu & 0 & 0 & 0 & -
\frac{\lambda_b}{\sqrt{2}} \,q \vspace{0.2cm} \\ 0 & \frac{
\lambda_t}{\sqrt{2}}\, t & \frac{g_2}{\sqrt{2}}\, q & 0 & 0 & 
\frac{\lambda_b}{\sqrt{2}}\, v_1 \vspace{0.2cm} \\ 0 & 0 & 0 & 
-\frac{\lambda_b}{\sqrt{2}}\, q & \frac{
\lambda_b}{\sqrt{2}}\, v_1 & 0 \end{pmatrix}
\label{eq:char}
\end{equation}
\item Neutralinos (mix between the fermionic partners of the neutral Higgses
$\tilde{H}_1^0$, $\tilde{H}_2^0$, the U(1) and SU(2) gauginos $\tilde{B}$, 
$\tilde{W}_3$, the colour index $3$ components of the top quark, 
$t_L^{(3)}$, $t_R^{(3)}$ and the eighth gluino $\tilde{G}^8$; 
$n_{1 \ldots 7} = -2$):
\begin{equation}
[M_{\chi^0}]\; =\;\,\begin{pmatrix} M_1 & 0 & -
\frac{g^\prime}{2}\, v_1 & \frac{g^\prime}{2}\, v_2 & -
\frac{g^\prime}{6}\, q & \frac{2 g^\prime}{3}\, t 
& 0 \vspace{0.2cm} \\ 
0 & M_2 & \frac{g_2}{2}\, v_1 & -\frac{g_2}{2}\, 
v_2 & \frac{g_2}{2}\, q & 0 & 0 \vspace{0.2cm} \\ 
-\frac{g^\prime}{2}\, v_1 & \frac{g_2}{2}\, v_1 & 
0 & -\mu & 0 & 0 & 0 \vspace{0.2cm} \\ 
\frac{g^\prime}{2}\, v_2 & -\frac{g_2}{2}\, v_2 & 
-\mu & 0 & -\frac{\lambda_t}{\sqrt{2}}\, t & -
\frac{\lambda_t}{\sqrt{2}}\, q & 0 \vspace{0.2cm} \\ 
-\frac{g^\prime}{6}\, q & \frac{g_2}{2}\, q & 0 & 
-\frac{\lambda_t}{\sqrt{2}}\, t & 0 & \frac{
\lambda_t}{\sqrt{2}}\, v_2 & -\frac{g_3}{\sqrt{3}}\, q 
\vspace{0.2cm} \\ 
\frac{2 g^\prime}{3}\, t & 0 & 0 & -\frac{
\lambda_t}{\sqrt{2}}\, q & \frac{\lambda_t}{\sqrt{2}}\, v_2 & 0 
& \frac{g_3}{\sqrt{3}}\, t \vspace{0.2cm} \\ 
0 & 0 & 0 & 0 & -\frac{g_3}{\sqrt{3}}\, q & 
\frac{g_3}{\sqrt{3}}\, t & M_3 \end{pmatrix}
\label{eq:neu}
\end{equation}
\item Gluinos (mix between the colour index $\{1 \,,\, 2\}$ components of the
top quark and the gluinos $\tilde{G}^{4\ldots 7}$; $n_{1\ldots 4} = -2 \times 
2$~\footnote{$t_L^{(1)}$, $t_R^{(1)}$ mix with $\tilde{G}^+ = (
\tilde{G}^5 + i\tilde{G}^4)/\sqrt{2}$, $\tilde{G}^- = (\tilde{G}^5 -i\tilde{G}^4
)/\sqrt{2}$. An identical mixing occurs between $t^{(2)}$, $\tilde{G}^6$
and $\tilde{G}^7$ producing degenerate gluino masses - thus, the extra factor
of $2$ in the coefficients $n$.}):
\begin{equation}
[M_{\tilde{G}}]\; =\;\,\begin{pmatrix} 0 & \frac{
\lambda_t}{\sqrt{2}}\, v_2 & -\frac{g_3}{\sqrt{2}}\, q & 0 
\vspace{0.2cm} \\ 
\frac{\lambda_t}{\sqrt{2}}\, v_2 & 0 & 0 & 
\frac{g_3}{\sqrt{2}}\, t \vspace{0.2cm} \\ 
-\frac{g_3}{\sqrt{2}}\, q & 0 & 0 & M_3 \vspace{0.2cm} \\ 
0 & \frac{g_3}{\sqrt{2}}\, t & M_3 & 0 \end{pmatrix}
\label{eq:gli}
\end{equation}
\end{itemize}
The remaining gluinos ($\tilde{G}^{1\ldots 3}$) contribute to the one-loop
potential with mass $M_3$ and are affected by a factor $n=-2$ each. At the 
tree-level minimum the matrices~\eqref{eq:mstop} and~\eqref{eq:mch} have each a 
zero eigenvalue, the matrix~\eqref{eq:mps} has two. Counting the multiplicities 
of each particle this corresponds to a total of eight Goldstone bosons, 
corresponding to eight gauge bosons that acquire mass (the $Z^0$, the $W^\pm$ 
and five of the gluons). We checked these mass matrices by computing $Str\,M^2 
\,=\, \sum_\alpha \, n_\alpha\,M_\alpha^2$ - because supersymmetry is broken in
a ``soft" way this quantity should be field-independent (that is, it should be
independent of the value of the vevs), and it is simple to verify that condition
is met.

\section{Minimisation of the one-loop CCB potential}
\label{sec:min}

With the mass particles for the whole sparticle spectrum obtained it is a 
simple task to obtain the derivatives of eq.~\eqref{eq:der} and perform the 
one-loop minimisation of the CCB potential as was done in ref.~\cite{min}. 
Here we will use a different method to minimise the 
potential: we will not bother with its derivatives and instead look directly
at the values of $V_0 + \Delta V_1$ as function of $(v_1 , v_2 , q, t)$ by means
of a modified simulated annealing algorithm incorporating Q-sampling of the 
phase space~\cite{jorge}. The code for the application of this algorithm was 
developed by J.M. Pacheco. The reason for choosing this method is purely a 
practical one: the algorithm is extremely efficient and the computation time 
necessary for a scan of the MSSM parameter space drastically reduced. We checked
the code by comparing its results with those of a MSSM calculation using the 
potential's derivatives, for a vast region of parameters. We also compared its
results with those obtained using the numerical minimisation tools of the MATLAB
package. All these approaches produced the same results. 

In this work we will mostly study the MSSM with universal input soft parameters.
At the gauge unification scale we choose common values $A_G$, $M_G$ and $m_G$ 
for the $A$ parameters, the soft gaugino and scalar masses. We took $-4 \leq A_G
\leq 4$ TeV and $20 \leq M_G, m_G \leq 900$ GeV.  Further, we have taken $2.5 
\leq \tan\beta \leq 10.5$ and considered both possible signs for the $\mu$ 
parameter. This selection of parameter space is by no means an exhaustive one 
(we left out high values of $\tan\beta$, for instance) but is already very 
extensive and should provide a good idea of the importance of the CCB bounds 
derived from our one-loop potential. We follow the ``top-bottom" approach 
outlined in ref.~\cite{bbo}: at the weak scale $M_Z$ we input $M_Z = 91.19$ GeV,
$M_t = 
167.2$ GeV, $M_b = 2.95$ GeV, $M_\tau = 1.75$ GeV (these are running fermion 
masses, not the pole ones), $\alpha_1 = 0.01667$, $\alpha_2 = 0.032$, $\alpha_3 
= 0.1$ and the value of $\tan\beta$.  Because the gauge and Yukawa 
$\beta$-functions do not depend on the soft parameters~\footnote{Except 
indirectly, in the form of particle thresholds.}, we can run these parameters up
in the energy scale until we find the gauge unification scale, defined as the 
value $M_U$ for which the couplings $\alpha_1$ and $\alpha_2$ meet. At that 
point we input the values of the soft parameters, chosen as explained above, and
run the whole theory to a scale $M_C = \mbox{max}(M_Z \,,\,M_G\,,\,g_3\,A_G/
\lambda_t)$ - this scale is a good estimate of the heaviest masses in the 
theory, thereby reducing, in principle, the size of the logarithmic 
contributions to the potential. At that scale we use the one-loop MSSM 
minimisation conditions (see, for instance, ref.~\cite{bbo}) and determine - if 
possible - the parameters $\mu$ and $B$. We then calculate the sparticles' 
masses~\footnote{With the tree-level mass matrices except for the neutral 
CP-even Higgses, for which full one-loop expressions were used.} and use recent 
experimental bounds~\cite{pdg} to reject those ``points" already in 
contradiction with observational evidence. We end up with over 39000 ``points" 
of parameter space. We also studied a small non-universal parameter space where
we took the universal ``points" obtained earlier and set the $m^2_t$ soft 
parameter to negative values (of the order of, at the most, $(100\; 
\mbox{GeV})^2$),
as this situation would seem one of the likeliest to result in CCB. This 
situation clearly requires non-universality, as universal values of the soft 
masses are unlikely to result in negative $m^2_t$ at the weak scale, due to the
form of the $\beta$-function for this soft parameter. With the new value of 
$m^2_t$ we minimised the (new) potential and, again, used experimental 
sparticles' mass bounds to reject ``points" in disagreement with observations. 
Our non-universal parameter space ended up with about 4000 ``points". 

We set out to determine the impact of the one-loop contributions on CCB bounds 
and as such we endeavored to compare results coming from tree-level and 
one-loop minimisations of the potential. The tree-level minimisation of the MSSM
potential is performed analytically in the standard way, see for 
example~\cite{bbo}. So, both at tree-level and one-loop, we compute the value of
the MSSM potential, $V^{MSSM}$, and perform the numerical minimisation of the 
CCB one, obtaining the vevs $\{v_1\,,\,v_2\,,\,q\,,\,t\}$. With these we 
calculate the value of $V^{CCB}$ and compare it with $V^{MSSM}$. Finally, a word
on thresholds: we follow the procedure of ref.~\cite{chan}, using the full MSSM 
$\beta$-functions from $M_Z$ to $M_U$, and choosing the input parameters at 
$M_Z$ such that the threshold contributions are automatically taken into account
- this is an effective procedure, but shown to produce good results. And for our
purposes - determining if CCB occurs or not - this degree of precision should be
more than adequate.   Being based on a Monte Carlo method the algorithm depends 
on the initial conditions used, so several runs of the program were necessary. 
We only accepted those extrema with all CCB squared masses positive, except the 
expected four zero eigenvalues corresponding to the Goldstone bosons. We 
remember that when performing a one-loop minimisation of the MSSM potential it 
is usual to find negative squared masses in the Higgs sector~\cite{gam}. They 
correspond to the Goldstone bosons which have zero masses for a tree-level 
minimisation. Since we do a one-loop minimisation and compute their masses using
tree-level matrices, negatives do occur sometimes. However, the absolute value 
of these negative squared masses is very small when compared to the other masses
in the theory and thus, based on~\cite{raif}, they can be safely set to zero. 

The results of this scan of the MSSM parameter space may be resumed as follows:
\begin{itemize}
\item No unbound-from-below (UFB) directions were found for the one-loop 
minimisations. They appear frequently, however, for the tree-level 
minimisations. It is easy to see from equation~\eqref{eq:vc}, for instance,
that with $q$ and $t$ constant and $v_1 \simeq v_2 \rightarrow 
\infty$ we have a potential tree-level UFB direction. 
These directions are characterized by vevs assuming arbitrarily large values, 
causing the potential to become arbitrarily negative. No such thing happened in 
the one-loop minimisation of the potential, which confirms the expectations of 
the authors of~\cite{cas}: they had assumed that the UFB directions they had 
found at tree-level might not be present if a one-loop minimisation was 
performed. We confirmed that that is the case, and that the one-loop 
contributions to the potential stabilize the vevs (which is also in agreement 
with the conclusions of~\cite{gam}). 
\item For the tree-level minimisations, the value of the CCB potential was found
to be {\em always} deeper than the MSSM one. For many of these minima, however,
the values of $q$ and $t$ found were practically zero. What happens is that the 
tree-level MSSM potential is found for very specific values of the vevs (namely 
such that $v_1^2 + v_2^2 = (246\;\mbox{GeV})^2$ and $\tan \beta = v_2/v_1$), but
that is not the only possible MSSM minimum - the tree-level MSSM minimisation
conditions consist of two coupled cubic equations and as such can have as much 
as nine different solutions. What we find is thati, for the CCB case, we let the
four vevs ``roam freely" as we minimise numerically the potential and, as such, 
many solutions with $q \simeq t \simeq 0$ are found. However, those solutions 
have values of $v_1$ and $v_2$ which are not phenomenologically acceptable 
(giving as they do wrong values for the gauge bosons' masses, for instance). 
These alternative minima, which preserve the MSSM gauge symmetry, are actually 
deeper than the ``standard" minimum. But again we must remember that a 
comparison of tree-level potentials with tree-level derived vevs is inherently 
flawed, and we only undertook it to show the radical differences with the 
one-loop case.
\item The MSSM potential is always deeper than the CCB one at one-loop, both for
the universal and non-universal parameter spaces. There are now no ``alternative
MSSM minima" as in the tree-level case, and the reason is easy to understand: 
while a given combination of supersymmetric parameters might correspond to 
several possible combinations of vevs as minimisation solutions at tree-level, 
it is almost impossible that that happens for the one-loop potential, given its 
complexity. 
\end{itemize}

The comparison between these two final points reveals how different a one-loop
procedure can be from the tree-level one. But it also requires some explanation:
what is it about the one-loop CCB contributions that raises the value of the 
potential, always above its MSSM value? First, we have already seen the 
importance of a one-loop minimisation to stabilise the values of the vevs
(avoiding UFB directions and preventing the appearance of ``alternative MSSM
minima"). Secondly, the consequences of the different gauge symmetry breakings 
become apparent only at the level of the one-loop potential, when we consider
the different spectra of masses thereof resulting. In section~\ref{sec:mas} we
showed that a very significant difference between the MSSM and the CCB case is
the mass of the gluons: all zero for the MSSM, five of them gaining mass for
CCB. For CCB vevs $q$ and $t$ of the same order of the electroweak vevs $v_1$ 
and $v_2$ (as we expect them to be, if not even higher), we expect the gluons to
have masses considerably larger than $M_Z$ or $M_W$ since, from 
eq.~\eqref{eq:mgl}, their masses are proportional to the strong coupling 
constant. As these are bosonic masses, their contributions to $\Delta V_1$, from
eq.~\eqref{eq:dv1}, should be positive. Assuming a CCB minimum exists, 
then, its gluon one-loop contribution to the potential is expected to be large
and positive, whereas for the same SUSY parameters the similar contribution to 
the MSSM potential is zero. This is a possible explanation, but the one-loop
contributions are very complex and the real reason may be lying with other 
terms. 

Ultimately, all we can conclude from these results is that we have a 
``numerical demonstration", for this vast parameter space, of the MSSM minimum
being deeper than any possible CCB minimum. This follows similar conclusions 
reached in ref.~\cite{min}, where no acceptable CCB minima associated with the 
tau Yukawa coupling were found. Also, recent work~\cite{bar} demonstrated that 
for two Higgs doublet models (2HDM) at tree-level, if a minimum preserving 
$U(1)_{em}$ and CP symmetries exists, it is a {\em global} minimum. In such 
models it is impossible to have tunneling for charge or CP breaking vacua. Of 
course, in this work we are dealing with charge and colour breaking, in a model 
that is far more complex than the 2HDM. We are also studying one-loop minima, 
not tree-level ones. But these different elements may be suggesting that the 
occurrence of CCB is not as easy to occur as a tree-level analysis in the MSSM 
leads us to believe. 

It is important to remark that we did not exhaust the MSMM parameter space. 
The possibility of dangerous CCB minima occurring may happen in some portion
of parameter space not included in our study. Even though the range and number 
of parameters we chose was very general, this possibility cannot be wholly  
dismissed. Finally, all our calculations have assumed that the MSSM vacuum is 
the absolute minimum of the theory. There is however~\cite{eve} the theoretical 
possibility that the the CCB minimum is deeper than the real one and the 
tunneling time between both vacua be superior to the age of the universe. This 
leads to a relaxation of the CCB bounds usually obtained. In our case we do not 
need to worry about this possibility since no CCB absolute minima were found. 

We hope to have convinced the reader of the importance of full one-loop 
calculations in estimating CCB bounds. In references~\cite{ccb},~\cite{min} and
this work we showed that for RG consistency one had to compare the one-loop 
MSSM and CCB potentials, and that the results of that enterprise gave very
different results from tree-level studies. We studied very specific CCB 
directions and vast sections of the MSSM parameter space, and no CCB minima
deeper than the ``normal" vacuum were found. Although at this stage we cannot
make sweeping generalisations and claim all CCB bounds used in the literature
are wrong, these results urge some caution. In reference~\cite{phe}, for 
example, large sections of parameter space are excluded on CCB grounds, and
predictions of supersymmetric masses are affected by it. If indeed, 
as this work suggests, tree-level CCB bounds are over-estimated, we may be 
excluding areas of parameter space that can be of experimental 
interest. A careful re-evaluation of CCB bounds might be in order. 

\vspace{0.25cm}
{\bf Acknowledgments:} My deepest thanks to Jorge Pacheco for providing me with
the code for the minimisation algorithm. This work was supported by a fellowship
from Funda\c{c}\~ao para a Ci\^encia e Tecnologia, SFRH/BPD/5575/2001.


\begin{thebibliography}{99}
\bibitem{fre} J.M. Fr\'ere, D.R.T. Jones and S. Raby, {\em Nucl. Phys.} {\bf 
B222} (1983) 11.
\bibitem{eve} L. Alvarez-Gaum\'e, J. Polchinski and M. Wise, {\em Nucl. Phys.}
{\bf B221} (1983) 495; 

J.P. Derendinger and C.A. Savoy {\em Nucl. Phys.} {\bf B237} 307; 

C. Kounnas, A.B. Lahanas, D.V. Nanopoulos and M. Quir\'os, {\em Nucl. Phys.} 
{\bf B236} (1984) 438; 

M. Claudson, L.J. Hall and I. Hinchliffe, {\em Nucl. Phys.} {\bf B228} (1983) 
501; 

M. Drees, M. Gl\"uck and K. Grassie, {\em Phys. Lett.} {\bf B157} (1985) 164;

J.F. Gunion, H.E. Haber and M. Sher, {\em Nucl. Phys.} {\bf B331} (1988) 320.
\bibitem{gam} G. Gamberini, G. Ridolfi and F. Zwirner, {\em Nucl. Phys.} {\bf
B331} (1990) 331.
\bibitem{cas} J.A. Casas, A. Lleyda and C. Mu\~noz, {\em Nucl. Phys.} {\bf
B471} (1996) 3.
\bibitem{phe} U. Ellwanger and C. Hugonie, {\bf hep-ph/9811386}; 

S. Abel and T. Falk, {\em Phys. Lett.} {\bf B444} (1998) 427; 

S. Abel and C. Savoy, {\em Phys. Lett.} {\bf B444} (1998) 119; 

S. Abel and B. Allanach, {\em Phys. Lett.} {\bf B431} (1998) 339.

OPAL Collaboration, {\em Eur. Phys. Jour} {\bf C7} (1999) 407; {\em ibid.}, 
{\em Eur. Phys. Jour} {\bf C12} (2000) 567.
\bibitem{one} H. Baer, M. Brhlik and D. Casta\~no, {\em Phys. Rev.} {\bf D54}
(1996) 6944.
\bibitem{ccb} P.M. Ferreira, {\em Phys. Lett.} {\bf B509} (2001) 120; {\em Err.}
{\em Phys. Lett.} {\bf B518} (2001) 333.
\bibitem{min} P.M. Ferreira, {\em Phys. Lett.} {\bf B512} (2001) 379; {\em Err.}
{\em Phys. Lett.} {\bf B518} (2001) 334.
\bibitem{bbo} V. Barger, M.S. Berger and P. Ohmann, {\em Phys. Rev.} {\bf D49}  
(1994) 4908.
\bibitem{ford} C. Ford, D.R.T. Jones, P.W. Stephenson and M.B. Einhorn, {\em 
Nucl. Phys.} {\bf B395} (1993) 17.
\bibitem{icq} P.M. Ferreira, {\bf hep-ph/0210024}.
\bibitem{ros} J. Rosiek, {\em Phys. Rev.} {\bf D41} (1990) 3464.
\bibitem{jorge} A.K. Hartmann and H. Rieger, {\em Optimization algorithms in 
physics}, Wiley VCH, Berlin 2002. 
\bibitem{pdg} K. Hagiwara {\em et al}, {\em Phys. Rev.} {\bf D66} (2002) 010001.
\bibitem{3loop} P.M. Ferreira, I. Jack and D.R.T. Jones, {\em Phys. Lett.} 
{\bf B387} (1996) 80.
\bibitem{pier} D. Pierce, {\bf hep-ph/9407202}.
\bibitem{chan} P.H Chankowski, Z. Pluciennik and S. Pokorski, {\em Nucl. Phys.} 
{\bf B439} (1995) 23. 
\bibitem{raif} Y. Fujimoto, L. O'Raifeartaigh and G. Parravicini, {\em Nucl.
Phys.} {\bf B212} (1983) 268.

E.J. Weinberg and A. Wu, {\em Phys. Rev.} {\bf D36} (1987) 2474.
\bibitem{bar} P.M. Ferreira, R. Santos and A. Barroso, {\em Phys. Lett.} {\bf 
B603} (2004) 219.
\end{thebibliography}
\end{document}